# An empirical study on a learning path on wave physics focused on energy

*Vera Montalbano and  Simone Di Renzone*
*Department of Physics, University of Siena, Italy*

Abstract

We describe an extracurricular learning path on waves focused on energy transfer. The advantages of introducing mechanical waves by using the Shive wave machine and laboratory activities are presented. Laboratories are realized by inquiry, i.e. students explore waves behavior in qualitative way, guess what can happen and suddenly test their hypothesis. Recently, we presented some disciplinary knots that arise usually in empirical investigation, according to the Model of Educational Reconstruction and discussed methodological choices made in designing the learning path and preliminary result about its realization with few, interested and talented pupils. We report the second year of this learning path performed with the same students that are introduced to more complex topics such as analogy in wave phenomena and resonance. Laboratories are described with particular attention for the energy transformation. We planned activities by focusing on conceptual issues such as characterization of the oscillatory motion and energy aspects vs. characterization of wave energy and energy transport. We designed the activities in order to propose a complementary experience compared to what was done in class. Despite resonance is a relevant phenomenon which runs through almost every branch of physics, many students have never studied it. Yet, resonance is one of the most striking and unexpected phenomenon in all physics and it easy to observe but difficult to understand. Students performed activities in laboratory on several resonant systems. Our purpose was to outline how it is possible to tune a system or a device in order to obtain resonance and an efficient energy transfer from different physical systems, such as a mechanical one and an electrical one. This year, the final task was to analyze different natural phenomena in order of choosing one suitable for energy transfer. We present our considerations on the students' learning process and on the possibility of extend a similar path in a classroom.

The description of wave phenomena in physics involves fundamental concepts that are difficult for many students (e. g. Wittmann, 1998; Wittmann, Steinberg, & Redish, 2003; Ambrose, Heron, Vokos & McDermott, 1999). Understanding and using functions of two variables, distinguishing between medium properties and boundary conditions, recognizing consequences of local phenomena in extended systems are few examples. More advanced topics, such as analogy, superposition or resonance, can be introduced simply or only in this context. Energy transport is another relevant topic in wave physics. Despite being such an interesting topic for students considering the economic, social, environmental and technological implications, it is rarely discussed at school.

Recently, we began to design learning paths on wave physics with the purpose of improving the achievement of students in this strategic topic. Since we used a methodology and laboratory activities quite different from standard teaching in our context, we designed an optional laboratory for performing a pilot investigation with students in secondary school within National Plan for Science Degree (Piano nazionale per le Lauree Scientifiche, i.e. PLS). Testing these activities with high school students can be considered the first step in order to develop a designed-based research learning path (Jonassen, Cernusca, & Ionas, 2007; Hake, 2008; Ruthven, Laborde, Leach, & Tiberghien, 2009).

The main aim was orienting toward physics in a more effective way, by introducing interested and talented students in wave phenomena through a most insightfully path. According to the Model of Educational Reconstruction (Duit, Komerek, & Wilbers, 1997; Duit, 2007), we identified some disciplinary knots that arise usually in empirical investigation (Di Renzone, Frati, & Montalbano, 2011).

Thus, we planned two extracurricular learning paths on waves, by following the guidelines of PLS. In the next section, we reported the methods and details of the investigation that was designed with a duration of 3 years. In the following section, we describe some relevant aspect and findings, in particular for the second year of the investigation. Finally, in the last section we discuss the preliminary results obtained until now.

## Learning Paths on Waves

### Overview

In recent decades it has been detected almost everywhere a consistent decrease of graduates in science disciplines. The situation in Italy was dreadful. Thus, the Ministry of Education and Scientific Research promoted a wide project in order to reverse this trend: National Plan for Science Degree (for a survey on PLS and all actions realized locally in this context see Montalbano, 2012). A relevant role in the plan is played by PLS laboratories. Our test with students were realized as optional extracurricular PLS laboratories while all the design activities of learning paths were developed and implemented within two courses of the Master in Physics Educational Innovation and Orienting (a PLS action in professional development for teachers, for more details see Montalbano, Benedetti, Mariotti, Mariotti, & Porri, 2012) in which one author is enrolled (S. D. R.). Our proposal consisted in two deepening laboratories for selected students titled Waves and energy and Sound and surroundings.

### Methods

The design was focused on conceptual issues, such as characterization of the oscillatory motion, wave energy and energy transport, and methodological issues in order to propose a complementary experience compared to what was done in class. Students ended their learning path on waves and sounds in class before laboratories started. Moreover, their class made an instruction trip to our department and perform a standard laboratory experience on diffraction and interference with light.

Both learning paths followed closely a type of PLS laboratory: Deepening laboratories for motivated and talented students. According to PLS strategy, they were orienting to science by means of training, students were the main character of learning and laboratory was thought as a method not as a place. Moreover, all activities arose from joint planning by teachers and university.

Since in previous PLS activities, we found a real effectiveness in active and cooperative behavior of students in laboratory (Benedetti, Mariotti, Montalbano, & Porri, 2011), we organized all meetings with a short introductory discussion followed by an experimental session in which students could explore waves behavior in qualitative way, guessing what could happen and suddenly testing their hypothesis. The realization by inquiry was utilized every time it was possible even when quantitative measurements were requested.

For evaluating the effectiveness of learning process we utilized direct observations in laboratories, resuming reports on measures requested to students at the end of relevant activities, annual final reports that students gave to their physics teacher for assessment of PLS laboratory in the school.

The laboratories were optional and the activities took place in Physics Department. We planned to meet students for 3 hours almost every month and to continue for the last 3 years of high school for a total every year of about 15-18 hr. We decided that for the first year both laboratories had the same introductory activities on waves physics and all students worked together (for a survey of activity in the first year, see Di Renzone, Frati, & Montalbano, 2011). In the second year, we decided that some other activity can be discussed by the two groups of students together or performing the same experiment with different tasks. In particular, this happened for the activity in which resonance is introduced and for all activity and discussion about similarities in physics. Thus almost an half of meetings were still joined at least in the introductory part.

**Waves and Oscillations**

**Activity description.** Waves were characterized in laboratory by using a Shive wave machine (described in details in the next section). Students were free of exploring and manipulating the device for having a prompt qualitative overview of phenomena. Then, some hint was given for obtaining quantitative information by using a camcorder. Students used Shive wave machine for studying: Wave dependence on space and time, Impulsive and periodic waves, Longitudinal and transverse waves, Wavelength and frequency, Energy transfer, Speed of propagation, Superposition principle, Reflection and transmission, and Energy conservation.

Sound waves were studied by using a microphone and an oscilloscope as an example of longitudinal waves. Students verified the principle of superimposition in this case and studied beats and patterns of periodic beats (Moiré fringes). Interference was studied for sound waves in order to stimulate reflection around similarities and differences between different kinds of waves (longitudinal vs. transverse, three-dimensional vs. two-dimensional, and so on).

In the second year, resonance was introduced through a mechanical system, a magnet suspended from a spring which can be forced by induction with an electromagnet. Students studied this system and others focusing on energy transfer, in the case of conservation as well as dissipation. Some other system, such as a vibrating string and a RLC circuit, were analyzed in order to clarify the concept of natural frequency of a system and how can be changed by chancing some system property. At this point, we arrived to introduce spectral decomposition of a broad signal and an example was given on dependence from boundary condition (vibrating string and resonant acoustic cavities).

Spectral analysis of solar white light, of a spectral lamp, of a laser beam and of a human sound comparison allowed students to start considering similarities in physics.

The final part of the second year was dedicated to considering energy transfer between mechanical or electromagnetic devices and natural phenomena. A research was made about which natural phenomena can be used as a suitable source of renewable energy.

**Participants.** Three students of 3[th] class in Liceo Scientifico *Aldi* in Grosseto (starting age 15-16).

**Sounds and Surroundings**

**Activity description.** The activities were common with wave learning path until resonance and spectral decomposition, but specific aspects were outlined such as the relevance of RLC circuit in modern musical instruments, many examples of resonant acoustic cavities were given and sometimes studied in detail.

In order to clarify the difference between one and two-dimensional vibrating systems a specific activity was performed in which students studied resonance in the case of metal sheets of different shape and acoustic waves.

**Participants.** Three students attending 3[rd] class of a Scientific High School (Liceo Scientifico *Aldi* in Grosseto), starting age 15-16. In the second year another student joined belonging from the same class of other participants.

## Relevant Aspects on Designing and Findings

**Overview**

We designed the learning path in the way showed in previous section in order to focusing on the following conceptual knots in which the main difficulties in learning usually appear:

- Waves as function of several variables; this usually is an hidden trouble. Even brilliant students can use for long times functions of one and two variables without any real understanding of differences.

- Superposition principle is a fundamental concept and can clarify many phenomena in wave physics.

- Energy transport in order to distinguish waves from other periodic phenomena and comprehend many applications.

- Analogy in waves phenomena; difficulties are very common and reported (Podolefsky, & Finkelstein, 2007), especially in recognizing the same behavior in different context such as total reflection, diffraction, beats, interference and so on.

- Resonance; despite it is a relevant phenomenon which runs through almost every branch of physics, it is easy to observe but difficult to understand.

In the following, we want to describe advantages of introducing waves analysis by means of a Shive wave machine and the activities that, we believe, were more interesting and relevant for students learning process in this second year (for a summary of the first year relevant activities, learning problems and discussion, see Di Renzone, Frati, & Montalbano, 2011).

**Shive Wave Machine**

This device, showed in Figure 1, was developed by Dr John Shive at Bell Labs in '50 (Shive, 1959), and consists of a set of equally-spaced horizontal rods attached to a square wire spine. Displacing a rod on one of the ends will cause a wave to propagate across the machine.

Torsion waves of the core wire translate into transverse waves. Measures were obtained by using a camcorder and extracted from the captured images.

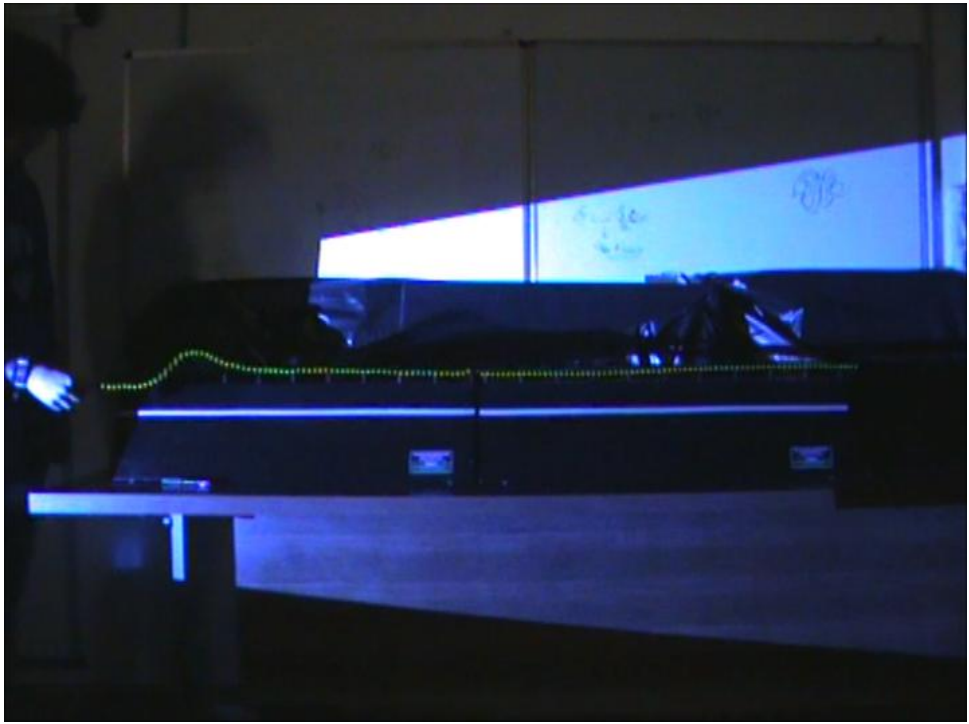

*Figure 1*. The Shive wave Machine is showed. A student has just perturbed an extremity for obtaining a pulse travelling from left to right in order to understand what happens when a wave encounter a discontinuity in the medium (left and right part of the device are connected and made by rods with different length). This figure is a frame from the camcorder; by analyzing different frame, students were able to measure the speed of waves in the two part of Shive wave machine. The blue light belongs from a video projector and was optimized by students for video recordings.

We choose Shive wave machine for characterizing mechanical wave because of easy and full interaction that allows to students. Measures of period, frequency, wavelength, speed of waves are straightforward; energy considerations, qualitative and quantitative tests are easy to perform. Stationary waves, reflection and resonance (or absence of it is) are simple to achieve and study. The main limit of the Shive wave machine is given by the fact that it produce a well-defined one-dimensional wave. Therefore, it is not possible to study refraction, diffraction and interference by using this device.

The enjoyment that students showed in using Shive wave machine and the forthcoming usage convinced us to utilize it also in the third year by constructing a tuneable system for transferring energy from it to an electric device.

**Similarities in Physics**

Shive Wave Machine was developed in order to point out similar features of waves as they propagate, reflect, superpose, resonate, etc. (Shive, 1959; Shive & Weber, 1982).

Analogies were displayed among the behaviours of waves on mechanical, acoustical, electrical, optical, electromagnetic systems every time that an activity in laboratory could be used for this purpose. Students were invited to consider different wave phenomena and find speed, frequency an wavelength range for:

- Mechanical waves: ripple tank, vibrating string, vibrating membrane, …
- Pressure waves: sound, ultrasound, ...
- Electromagnetic waves: radio waves, microwaves, light, infrared, ultraviolet, X rays, gamma rays, ...

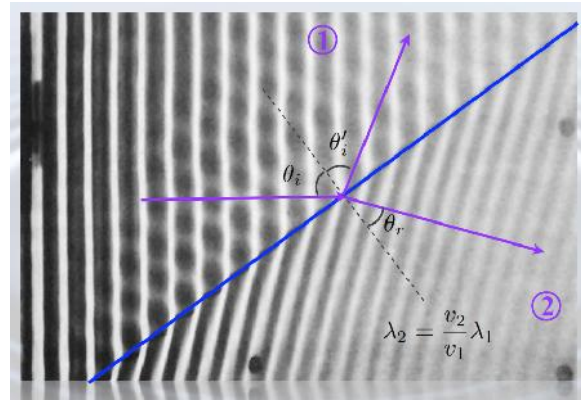

*Figure 2*. On recognizing Snell's law for waves on a liquid surface.

Then, they had the task of recognizing transverse and longitudinal waves, beats, reflection, refraction, interference, diffraction, Doppler effect in these different phenomena. In Figure 2 and 3 are shown two examples of refraction we gave, i.e. for a wave in liquid surface and for sound. We intentionally left students free to include other phenomena in each class if deemed appropriate and interesting. Thus, they proposed infrasound and seismic waves; an interesting discussion emerged from deepening the definition of radio waves.

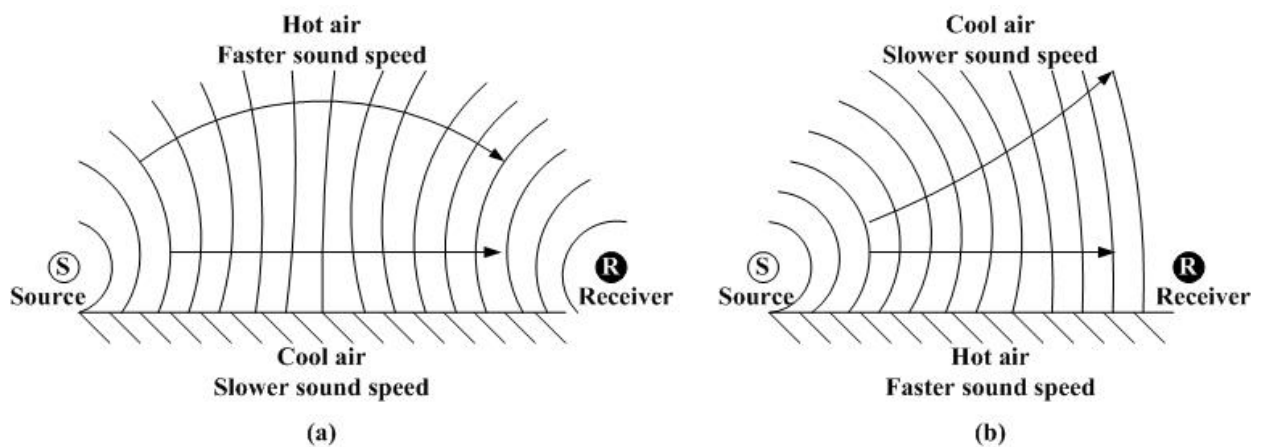

*Figure 3*. Sound refraction from temperature gradients (Sound refraction).

**Resonance**

In order to introduce students to resonant systems, we proposed a mechanical system, a magnet suspended from a spring which can be forced by induction by an electromagnet (a schematic set-up is given in Figure 4). Students studied this system focusing on energy.

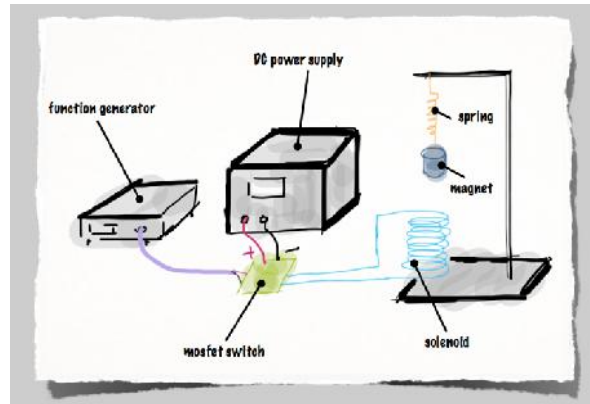

*Figure 4*. A schematic set-up for the resonant system. A ceramic magnet was suspended from a spring on an electromagnet. The driving force in the experimental setup was supplied by a solenoid. To allow its modulation, a mosfet transistor was used as an electronic switch. A function generator applied a square wave of given frequency to the mosfet gate, resulting in a modulation of the DC current provided by the power supply to the solenoid. The solenoid forced the magnet to oscillate at the same frequency of the square wave: changing the frequency is it possible to study the resonant condition of the magnet-spring system.

Electromagnetic induction can easily transform mechanical energy into electrical energy. Moving magnet induce a electromotive force in the solenoid that can move electrical charges (see Figure 5). Enveloping the outside of the electromagnet with aluminium is possible to observe the damping due to eddy currents.

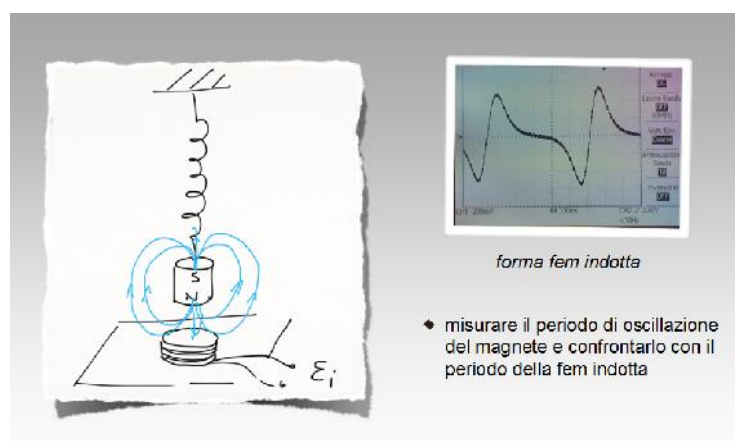

*Figure 5*. On the left a schematic set-up is showed. On the right, a picture of the display of an oscilloscope shows the electromotive force induced at the terminals of solenoid. Finally, a graphic obtained by a simple simulation allows to explain the form of the electromotive force in terms of velocity (green line) and distance from solenoid (red line) of the magnet.

Students studied the electromagnetic dumping in a quantitative way in the case of a pendulum in which the mass was a magnet, showed in Figure 6.

Can induction transform electrical energy in mechanical energy? Students discovered that this is not always possible. If they tried to modulate an electromagnet, only in few cases the energy transfer was massive. They changed frequencies in the system in Figure 4 and observed that only for one frequency there was a massive oscillation of the mechanical system. They measure the period of the oscillating system without electromagnet and compare the frequency with the one measured before discovering that it was the same. Only if the two systems, electromagnetic and mechanical, were oscillating at the same frequency they can exchange a large amount of energy. Since the mechanical system had only few natural frequencies this happened only in few cases. Students were able to discover another frequency, more than the one of spring-magnet systems, which corresponded to an oscillation like a pendulum of the magnet.

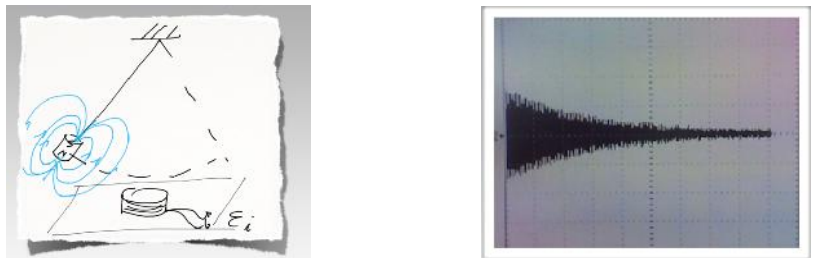

*Figure 6.* A schematic set-up of experiment of pendulum dumping is showed on the left. A ceramic magnet was suspended on a solenoid. Students visualized the electrical tension at the terminal of solenoid on the display of an oscilloscope. Copper plates were placed on the electromagnet for studying damping vs. thickness of the metal. On the right a picture of the display of oscilloscope shows the exponential dumping.

Students in the learning path on sounds performed another experience for resonance. By using a speaker connected to a function generator, a resonant system could be obtained by placing a metal plate over it. When sound was resonant with one frequency of the plate, salt started jumping leaving from the vibrating surface and cumulating in fixed zones. The figures formed by salt (Chladni figures, see figure 7 for two example) depend on the shape of plate, material, thickness and boundary conditions (existence of constrained points). In this case is very easy to recognize resonance.

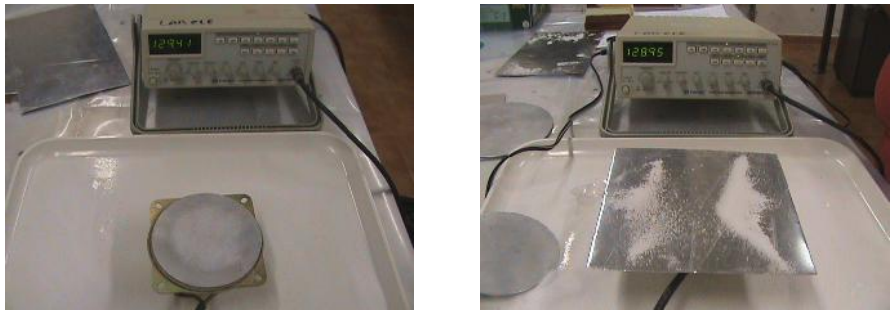

*Figure 7.* On the left a Chlandi figure is showed for a round sheet. On the right, a different picture for a rectangular sheet is given.

## Discussion and Conclusions

These laboratories seemed to be very successful, but sometimes students showed some learning difficult, in collecting properly and in using correctly the experimental data in order to describe physical systems, because they are little accustomed to face open situations. Students were very smart in using new device or in simple lab task such as measure of a period, but they showed a certain naivety in dealing with open problems.

Moreover, they were very active in laboratory but we obtained few elaborated materials. Every time we gave a task to complete by themselves at home, it was difficult to have back reports in a reasonable time and usually they were few accurate. We believe that the fact that their teacher was present only for reading the final report is the reason of this lack of care in activities performed at home.

The few involvement of the physics teacher implies also that all observations that we made in the students' response and all suggestions that we can give for improve and integrate this learning path with the teaching process at school are postponed. In this way, it is very difficult to foresee a direct impact of this experience on the practice at school.

On the other side, reflections and conclusions that we can draw from the results obtained so far are in the direction that these learning process can be performed at secondary school, at the price of a hard work for teachers, but likely it would be more useful in a dedicated course for undergraduate students.

## Acknowledgement

This work is based on activities and experiences which were realized within the National Plan for Science Degree supported by Italian Ministry of Education, University and Research. The authors would like to thank Stefano Veronesi, Serena Frati and Luca Marmugi for useful discussion and support in some activity in laboratory.